\documentclass[preprintnumbers,amsmath,amssymb,apl,
twocolumn,superscriptaddress]{revtex4}

\usepackage{graphicx}   
\usepackage{dcolumn}    
\usepackage{bm}         

\begin{document}

\title{Graphene quantum dots formed by a spatial modulation of the Dirac gap}

\author{G. Giavaras}
\affiliation{Advanced Science Institute, RIKEN, Wako-shi, Saitama
351-0198, Japan}

\author{Franco Nori}
\affiliation{Advanced Science Institute, RIKEN, Wako-shi, Saitama
351-0198, Japan} \affiliation{Department of Physics, The
University of Michigan, Ann Arbor, MI 48109-1040, USA}



\begin{abstract}
An electrostatic quantum dot cannot be formed in monolayer
graphene, because of the Klein tunnelling. However, a dot can be
formed with the help of a uniform magnetic field. As shown here, a
spatial modulation of the Dirac gap leads to confined states with
discrete energy levels, thus defining a dot, without applying
external electric and magnetic fields. Gap-induced dot states can
coexist and couple with states introduced by an electrostatic
potential. This property allows the region in which the resulting
states are localized to be tuned with the potential.
\end{abstract}

\maketitle

The relativistic character of electrons in graphene has attracted
considerable attention~\cite{neto2009,abergel2010}. This work
presents a way of forming a graphene quantum dot as a consequence
of a spatial modulation of the Dirac gap.

In an ideal graphene sheet, the band structure has no energy gap
and the Dirac electrons are massless. The quantum states of a
graphene dot, formed by an external electrostatic potential, are
deconfined due to the Klein tunnelling that is inherent to
massless particles~\cite{matulis2008}. Therefore, an electrostatic
dot cannot confine electrons in graphene since these can tunnel
through any potential barrier. However, a uniform magnetic field
suppresses the Klein tunnelling leading to confined
states~\cite{giavaras2009,maksym2010}.

It is experimentally possible to engineer an energy gap in
graphene's band structure, referred to as a Dirac gap, with a
value ranging from a few to hundreds of
meV~\cite{zhou2007,enderlein2010,vitali2008}. The gap leads to
electrons with mass and thus an electrostatic potential results in
confined states as in common semiconductors~\cite{abergel2010}.
Most importantly, as shown here, a spatially-modulated gap induces
confined states regardless of the application of external fields.
This can be achieved provided the gap has a local minimum in which
the states become localized, thus defining a quantum dot. The
application of a potential, generated by a gate electrode, couples
the gap- and potential-induced states. The coupling strength is
tunable with the potential, and it determines the region in which
the resulting states are localized.

A spatially-modulated Dirac gap has been
reported~\cite{vitali2008}. Possible ways of creating the required
gap modulation that forms the quantum dot include substrate
engineering and the application of strain to the graphene
sheet~\cite{neto2009,abergel2010}.

The physics of a graphene dot, for energies near the Dirac points,
is described by the Hamiltonian~\cite{abergel2010}
\begin{equation}\label{dothamiltonian}
H=v_F\bm{\sigma}\cdot(\bm{\mathbf{p} +
e\mathbf{A}})+V\mathcal{I}+\tau\Delta\sigma_{z},
\end{equation}
where the Fermi velocity $v_F$=$\gamma/\hbar$, with $\gamma$=646
meV nm, is assumed position-independent.
$\bm{\sigma}$=$(\sigma_{x},\sigma_{y})$, $\sigma_{z}$ are the
$2\times 2$ Pauli operators acting on the two carbon sublattices,
$\mathbf{p}$=$-i\hbar\bm{\nabla}$=$-i\hbar(\partial_x,\partial_y)$
is the 2D momentum operator, $\mathbf{A}$ is the vector potential
that generates the magnetic field
$\mathbf{B}$=$\bm{\nabla}$$\times$$\mathbf{A}$, $V$ is the
electrostatic potential and $\mathcal{I}$ is the unit matrix. The
last term in Eq.~(\ref{dothamiltonian}), referred to as mass term,
gives rise to an energy gap $2\Delta$ in the spectrum of graphene,
where $\tau$=1 ($\tau$=$-1$) corresponds to the $K$ ($K'$) valley.

For the dot model, $V$ and $\Delta$ are chosen cylindrically
symmetric and the magnetic field is uniform and perpendicular to
the graphene sheet, $\mathbf{B}$=$B\hat{z}$, so
$\mathbf{A}$=$(0,A_{\theta},0)$, with $A_{\theta}$=$Br/2$. The
Dirac equation $H\Psi$=$E\Psi$, can be written in cylindrical
coordinates with $\Psi$=$r^{-1/2}\{f_{1}(r)\exp[i(m-1)\theta], i
f_{2}(r)\exp(i m\theta)\}$, where $m$=$0,\pm 1... $ is the angular
momentum quantum number. The radial functions $f_{1}$ and $f_{2}$
satisfy
\begin{eqnarray}
(V-E+\tau \Delta)f_{1} +\left(U+\gamma\frac{d}{dr} \right)f_{2} &=&0,\label{radialequations1}    \\
\left(U-\gamma\frac{d}{dr}\right)f_{1}+(V-E-\tau \Delta)f_{2}
&=&0,\label{radialequations2}
\end{eqnarray}
with $U$=$\gamma(2m-1)/2r+\gamma eBr/2\hbar$.
Equations~(\ref{radialequations1}) and~(\ref{radialequations2})
are satisfied for both confined and deconfined states. The former
have an exponential tail asymptotically, e.g., in the limit of
large radial distance $r$ ($r$$\rightarrow$$\infty$), whereas the
latter have an oscillatory tail. If for large $r$ $V$ and $\Delta$
are constant or have a power-law dependence, then the
confined-deconfined character of a state is determined by the
asymptotic sign of
\begin{equation}\label{k2dependence}
q(r)=-\left(
\frac{eBr}{2\hbar}\right)^{2}+\left(\frac{V-E}{\gamma}\right)^{2}-\left(\frac{\tau\Delta}{\gamma}\right)^2.
\end{equation}
A state with energy $E$ is confined only if $q$ is asymptotically
negative~\cite{giavaras2009}. Otherwise the state is deconfined.
This criterion indicates that confined states can be induced even
for $B$=0 and $V$=0 everywhere, provided that $E^2-\Delta^2$$<$0
asymptotically. This inequality cannot be satisfied when $\Delta$
is spatially-independent because all the energies satisfy
$|E|$$>$$\Delta$. But the inequality can be satisfied when
$\Delta$ is spatially-dependent with an asymptotic value larger
than that for small $r$. This happens, for example, when $\Delta$
is zero within a disc area, and nonzero outside that area,
$\Delta$=$\delta_0$. Then a number of discrete energy levels
satisfy $|E|$$<$$\delta_0$ and correspond to confined states with
a large amplitude within the disc area. These states can be
regarded as dot states.

In the presence of an electrostatic potential $V$ and $B$=0,
Eq.~(\ref{k2dependence}) shows that if $V$ and $\Delta$ are
unequal and rise asymptotically, then confined states occur only
if ($V-\Delta$$)<$0 so that $q$$<$0. In this case confinement is
energy independent. However, this work focuses on the case where
both $V$ and $\Delta$ are constant asymptotically, which is the
most common experimental regime. Then confinement occurs if
$(V-E)^2-\Delta^2$$<$0 and thus it is energy-dependent.

\begin{figure}
\includegraphics[width=7.1cm]{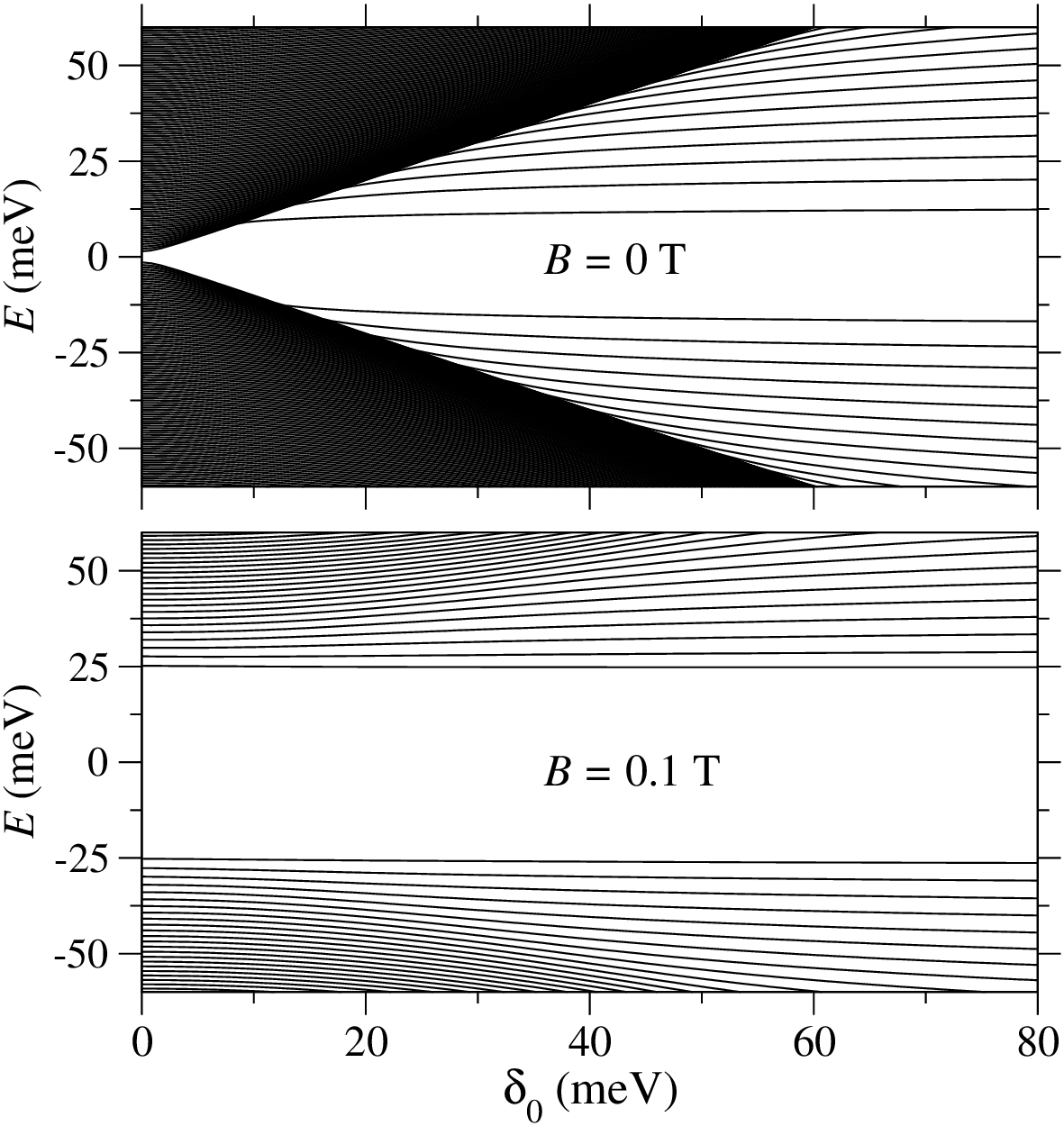}\\
\caption{Energy levels versus the asymptotic value of the mass
term $\delta_{0}$, that generates a Dirac gap
$2\delta_{0}.$}\label{vbzero}
\end{figure}

The properties of the gap-induced dot are analyzed by solving
numerically the two coupled equations
[Eqs.~(\ref{radialequations1}) and~(\ref{radialequations2})] using
a discretisation scheme, which satisfies time-reversal symmetry:
$E(m,B)$=$E(1-m,-B)$ for $\Delta$=0 and
$E(m,B,\tau)$=$E(1-m,-B,-\tau)$ for $\Delta$$\neq$0. The
spatially-dependent mass term is modelled by $\Delta$=0 for
$r$$\leq$$R$ and
$\Delta(r)$=$-\delta_{0}$/$\cosh[(r-R)/d]^{2}$+$\delta_{0}$ for
$r$$\geq$$R$ so that asymptotically $\Delta$$\approx$$\delta_0$.
This choice is not of particular importance; either a smooth or
sharp modulation of $\Delta$ results in confined states. For
brevity, all the results shown are for $R$=250 nm, $d$=150 nm,
$\tau$=1 and $m$=5.

Figure~\ref{vbzero} shows the energy levels as a function of the
asymptotic value of the mass term $\delta_0$ for $B$=0 and $V$=0.
Deconfined states correspond to the (quasi) continuum of levels
indicated by the black area in Fig.~\ref{vbzero} ~\cite{note1}.
Confined states correspond to the discrete levels emerging through
the continuum via anticrossing points (see below), forming two
distinct ladders of energy separated by a gap. The spacing of the
discrete levels and the gap increase with $\delta_0$. Further, the
confinement becomes stronger with $\delta_0$, and therefore an
increase in $\delta_0$ leads to an increase in the number of
discrete levels~\cite{giavaras2010}. However, for a fixed
$\delta_0$, not all $m$ values give confined states, since the
angular momentum term $U$, for $B$=0, tends to delocalize the
states for large $m$. For this reason there are no confined states
for $\delta_{0}$$\lesssim$10 meV in Fig.~\ref{vbzero}. The physics
is different if $B$$\neq$0 and $V$ is constant since from
Eq.~(\ref{k2dependence}) $q$$<$0 asymptotically, leading to
confined states independent of energy. One such case is
illustrated in Fig.~\ref{vbzero} for $B$=0.1 T. The continuum of
levels has been replaced by a discrete set for all $\delta_{0}$,
reflecting the disappearance of deconfined states. This is valid
for all $m$ values and $\tau$=$-$1.

\begin{figure}
\begin{centering}
\includegraphics[width=7.8cm]{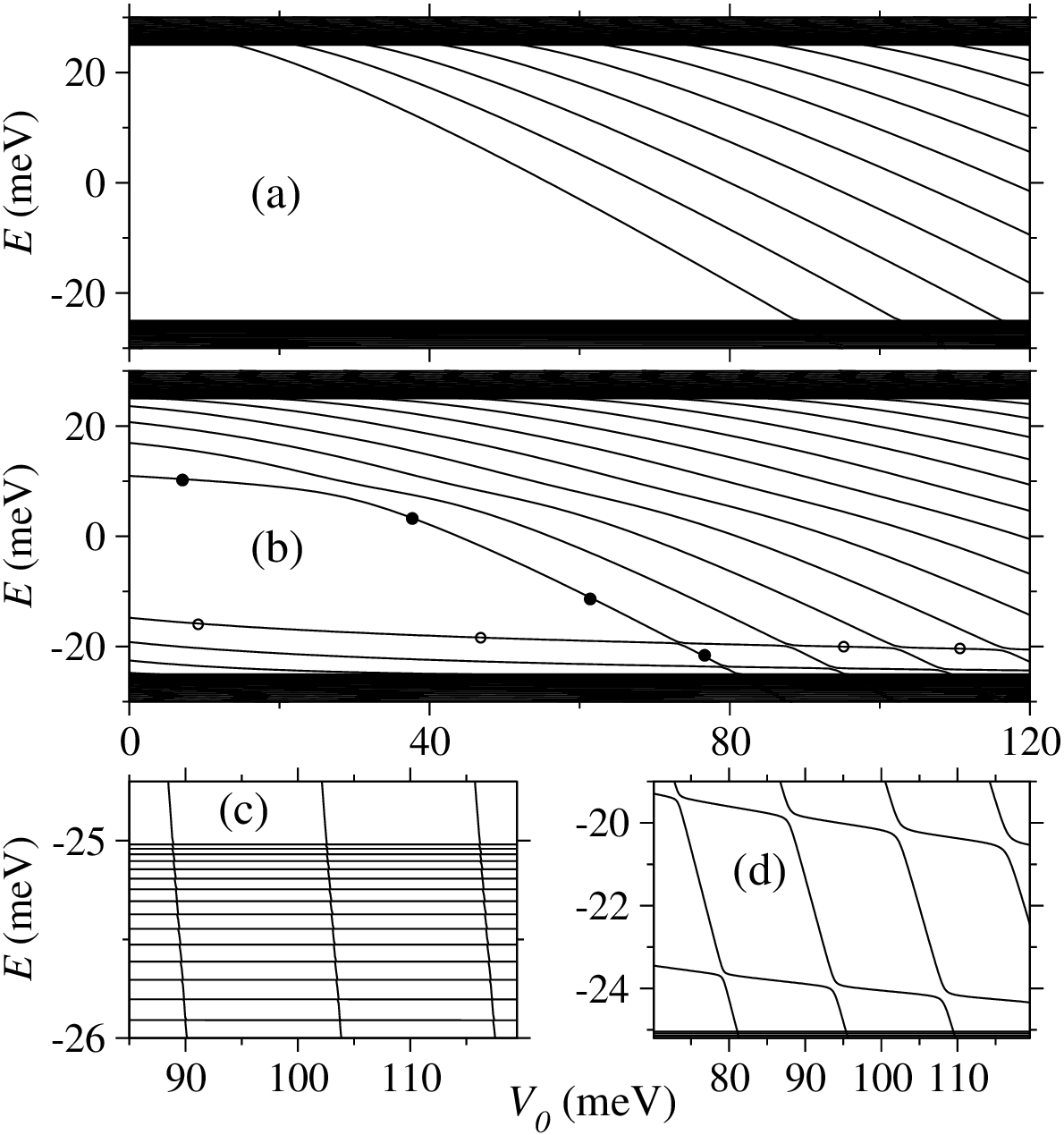}\\
\caption{(a) Energy levels versus the potential depth $V_{0}$ for
a constant Dirac gap $2\Delta$=50 meV. (b) As (a) but for a
spatially-modulated Dirac gap with an asymptotic value of
$2\delta_{0}$=50 meV. The states of the energies marked by circles
are shown in Fig.~\ref{states}. Enlarged views of (a) and (b) are
shown in (c) and (d).}\label{gap}
\end{centering}
\end{figure}

The effect of an electrostatic potential on the gap-induced dot is
now investigated. The potential that is generated in the graphene
sheet by gate electrodes can be calculated within the Thomas-Fermi
model~\cite{giavaras2009}. A slowly varying quantum well potential
is approximated by $V(r)$=$-V_{0}\exp(-r^2/l_0^{2})$. The quantum
well depth is $V_0$, the width is $l_{0}$, and asymptotically
$V$$\approx$0.

Figure~\ref{gap} shows the energy level diagram as a function of
$V_{0}$ for $l_{0}$=180 nm. Confined and deconfined states are
identified as in Fig.~\ref{vbzero}. For a constant gap and small
$V_{0}$, the angular momentum delocalizes the states; therefore
confined states are formed after a critical value of $V_{0}$. The
general trend is that with increasing $V_{0}$, the number of
discrete levels increases while the lowest levels merge into the
continuum. When this happens, the corresponding states undergo a
transition from confined to deconfined, which is reflected in the
energy diagram by the appearance of anticrossing points
[Fig.~\ref{gap}(c)]. These also appear when the states undergo the
opposite transition for energies near $\delta_{0}$. For a
spatially-dependent gap, there exist discrete levels even for
$V_{0}$=0, because of the gap-induced confinement. Unlike the
constant-gap system, as $V_{0}$ increases, anticrossing points are
formed between discrete levels [Fig.~\ref{gap}(d)], reflecting a
coupling between confined states due to the potential and the
spatially-dependent gap. This coupling is strong for the
gap-induced states of the upper ladder of energy and therefore the
corresponding anticrossing points are not well-defined. In
contrast, states of the lower ladder couple weakly to the
potential.

\begin{figure}
\includegraphics[width=8.1cm]{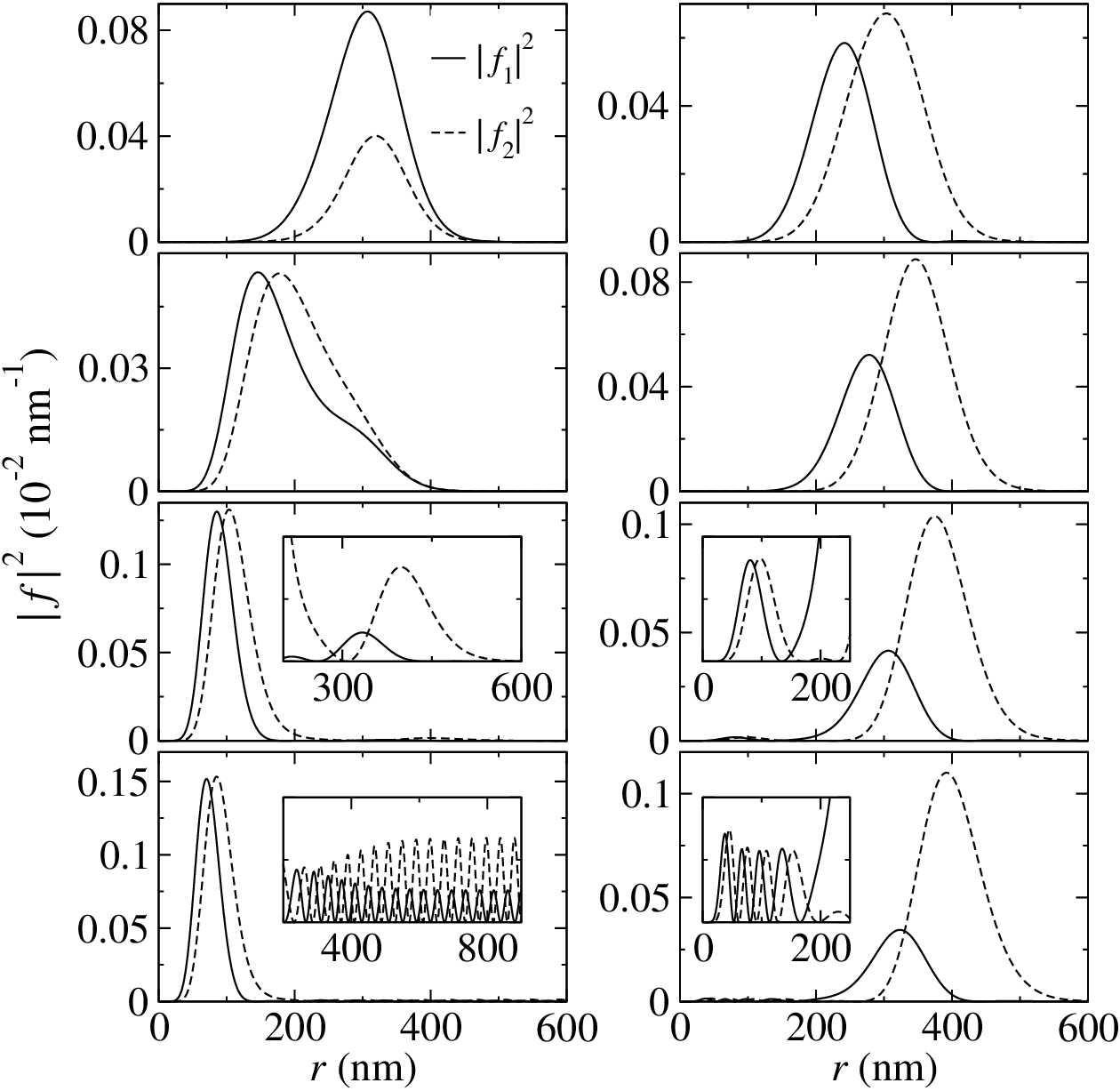}\\
\caption{Quantum states for different potential depths $V_{0}$,
from top to bottom: $V_{0}$=0, 30, 75, 120 meV. The Dirac gap is
spatially modulated with an asymptotic value of $2\delta_{0}$=50
meV. Left (right) panels show states with energies marked by
$\bullet$ ($\circ$) in Fig.~\ref{gap}(b). The vertical axes of the
insets range from 0 to 2$\times$10$^{-3}$.}\label{states}
\end{figure}

Figure~\ref{states} illustrates the effect of the potential on dot
states, for the energies shown in Fig.~\ref{gap}(b). For
$V_{0}$=0, the states are confined owing to the gap modulation.
Consider now the states shown in the left panels. As $V_{0}$
increases, the gap-induced state with positive energy couples to
the potential, e.g. for $V_{0}$=30 meV, and with increasing
$V_{0}$ it becomes localized in a region defined by the potential.
This state then couples to gap-induced states of the lower ladder,
e.g. for $V_{0}$=75 meV, while its energy decreases. For
$V_{0}$=120 meV the state is deconfined, with an oscillatory tail,
and its energy lies in the continuum. A coupling between gap- and
potential-induced states occurs also for the states shown in the
right panels, e.g. for $V_{0}$=75, 120 meV. As $V_{0}$ increases
the state with the maximum energy in the lower ladder (for
$V_{0}$=0), couples with excited potential states. This trend is
consistent with the series of anticrossing points in the
$E(V_{0})$ plot.

In Fig.~\ref{states} the coupling between potential- and
gap-induced states of the lower ladder is weak ($V_{0}$=75 meV);
therefore the states peak in the region defined either by the
potential profile (left) or the gap modulation
(right)~\cite{note2}. These two regions have a small overlap when
$R$$>$$l_{0}$ and $V_{0}$ is large. Strong coupling can be induced
for small $m$; for instance, for $m$=1 the states peak strongly in
both regions. The $m$-dependence of the coupling can be explained
within a semiclassical approach~\cite{abergel2010}. The relative
maximum amplitude of the two components, as can be derived from
Eqs.~(\ref{radialequations1}) and~(\ref{radialequations2}),
satisfies $|f_{1}|$$\sim$$|f_{2}|$ within a $\Delta$=0 region,
whereas $|f_{1}|$$>$$|f_{2}|$ ($|f_{1}|$$<$$|f_{2}|$) for energies
in the upper (lower) ladder. The latter behavior is more
pronounced when $\Delta$ is large and constant with
$V_{0}$$\neq$0. Then one of the components becomes vanishingly
small depending on choice of energy and valley ($\tau$=$\pm$1).

In order to probe the states of the gap-induced dot the Fermi
level has to be adjusted near the middle of the gap, where only
confined states with small values of $m$ lie, and hence the
resultant density of states is low. For the same reason the
electrostatic potential has to be small. Then, it should be
experimentally possible to resolve the quantum states using
similar measurements as in GaAs quantum dots.

In summary, a graphene dot can be formed as a result of a spatial
modulation of the Dirac gap, without applying external fields. An
electrostatic potential allows gap- and potential-induced states
to coexist and become coupled as the potential increases. The
coupling strength determines the region in which the states are
localized.

We thank P.A. Maksym and A.V. Rozhkov for discussions. G.G.
acknowledges support from JSPS. F.N. acknowledges support from
LPS, NSA, ARO, AFOSR, DARPA, NSF grant No.~0726909, JSPS-RFBR
contract No.~09-02-92114, Grant-in-Aid for Scientific Research
(S), MEXT Kakenhi on Quantum Cybernetics, and Funding Program for
Innovative R$\&$D on S$\&$T (FIRST).



\begin{thebibliography}{10}
\bibitem{note1}
The finite system size leads to a gap at $\delta_{0}$=0, though it
does not affect the discrete levels of the dot.
\bibitem{note2}
This latter region shifts at larger $r$ with $V_{0}$, and does not
bind for a too large $V_{0}$.
\end{thebibliography}
\end{document}